\documentclass{article}

\usepackage{graphicx}

\usepackage{amssymb,amsfonts,amsmath}

\begin{document}

\title{Change and Aging\\
Senescence as an adaptation}

\author{Andr\'e C. R. Martins\\
GRIFE -- Escola de Artes, Ci\^encias e Humanidades\\
Universidade de S\~ao Paulo -- Brazil \vspace{0.1cm}\\
 {\small \it also at} Centre de Recherche et Epistemologie Appliqu\'ee\\
\'Ecole Polytechnique -- France
}

\date{amartins@usp.br}

\maketitle

\begin{abstract}

Understanding why we age is a long-lived open problem in evolutionary biology. Aging is prejudicial to the individual and evolutionary forces should prevent it, but many species show signs of senescence as individuals age. Here, I will propose a model for aging based on assumptions that are compatible with evolutionary theory: i) competition is between individuals; ii) there is some degree of locality, so quite often competition will between parents and their progeny; iii) optimal conditions are not stationary, mutation helps each species to keep competitive. When conditions change, a senescent species can drive immortal competitors to extinction. This counter-intuitive result arises from the pruning caused by the death of elder individuals. When there is change and mutation, each generation is slightly better adapted to the new conditions, but some older individuals survive by random chance. Senescence can eliminate those from the genetic pool. Even though individual selection forces always win over group selection ones, it is not exactly the individual that is selected, but its lineage. While senescence damages the individuals and has an evolutionary cost, it has a benefit of its own. It allows each lineage to adapt faster to changing conditions. We age because the world changes.

\end{abstract}

\section{Introduction}
Living organisms shouldn't age, at least if that could be helped (many of use would certainly like that, but our wishes are not a valid argument). Evolution works in a way that any species whose representatives have any distinct disadvantage will be driven to extinction. It makes sense then to assume that, if aging could be avoided, species that showed senescence as the individuals grow older should be replaced by others where aging does not happen (or happens at a much slower rate). Senescence increases mortality and an individual who dies of old age will leave, in average, a smaller number of descendants than another individual that does not age and manages to live and reproduce for a longer time. And yet many known living organisms show senescence. The time it takes for an individual to show signs of old age varies greatly among species, but aging seems so natural that many people fail to realize there is an apparent contradiction between senescence and evolution.
Understanding the reasons why animals do age, despite or because of evolution, has important consequences in the prospect of whether aging is something we can avoid or not. If there is a mechanism for getting older, researchers can try to find ways to turn that mechanism off, with profound medical and demographical implications \cite{goldsmith08a,vijgcampisi08a}. 

General theories of aging are too many to be all listed here \cite{medvedev90a}. Many attempt to describe the mechanisms behind aging. However, while understanding the exact biological processes that lead to aging is fundamental, especially if we want to be able to counter some (or maybe, eventually, all) of its effects, it does leave one important question unanswered. That is the question of why. Since aging is so pervasive, even though there are organisms like the hydra that shows no signs of senescence \cite{martinez97a}, it seems reasonable to conclude that there might be general common principles behind it. The natural question becomes, from an evolutionary point of view, why most organisms have evolved to a situation where the individuals are, in some sense, harmed by some senescence mechanism \cite{rose91a}.

The absence of many elder individuals in natural conditions means that any detrimental effects will have lesser evolutionary importance if those effects only happen on advanced ages, since fewer individuals will be subject to it \cite{medawar52a}. If individuals, during their life time, will accumulate damage that would cause them problems, we should have aging, according to the accumulated mutations theory. However, the main question is not if animals suffer damage that might be accumulated, but if it is repairable. If organisms can avoid the accumulation by maintenance functions, we could, in principle, not get old. 

Antagonistic pleiotropy theory \cite{williams57a}, on the other hand, proposes that there might be genetic adaptations that, while providing benefits early in life, could be associated with problems later on. Since many more animals would receive the benefits, as rate of survival to old age is small, those benefits can compensate the posterior damage. Conditions for optimality for the theory have been studied \cite{sozouseymour04a}, showing  that, if some genes fit the description of the antagonistic pleiotropy hypothesis, there are circumstance where evolution will work in favor of keeping those genes. This does not mean, however, that most of the aging process is based on this effect.

The other evolutionary theory of aging that is based on individual selection is the disposable soma theory \cite{kirkwood88a,kirkwoodaustad00,kirkwood02,kirkwood08a}. The idea is that all animals need to have repair and maintenance systems to keep them functioning properly. These systems, according to disposable soma theory, need to be good enough to work during most of the reproductive life of an animal. However, since most animals in the wild die relatively young \cite{austad97}, there would be no need for assigning valuable resources to a repair system to keep it functioning longer. Optimal levels of maintenance would be dependent on the environment and species, thus explaining why different species age at very different rates. Evidence that species that are less subject to early accidental deaths live longer do exist and analysis of the life spans of social insects do seem to agree with the qualitative predictions of the theory \cite{kellergenoud97a}. 

These theories, while possible correct for several specific cases, are not capable, however, of explaining the whole range of observations related to aging. Accumulated mutation theory predicts that all organisms should show signs of senescence and that those organisms should be increasingly damaged the longer they live. However, not only some animals show no signs of aging, some, like female turtles \cite{congdonetal03a}, show evidence of increasing fertility with aging, in the contrary direction to what the theory predicts. Disposable some theory predicts that, if some mutation would increase the life span of an animal, it should decrease its metabolism, since resources would have to be diverted from other functions. However, genetic mutations that allow for longer life and cause no other losses have been observed \cite{kenyonetal93,arantesoliveiraetal03a,ayyadevara07a}. Molecular genetic studies have also shown that longevity can be subject to regulation and interventions can even reverse aging in animals \cite{ljubuncicreznick09a}. At this point, it seems that there is a growing body of evidence that aging might not be completely connected to fertility. This means that it is possible for animals to live longer without costs to their early life fertility \cite{mitteldorf04a}.

We are led to conclude that senescence might actually be an adaptation by itself, that it might not be a detrimental consequence of other gains. If that were the case, it would have to cause benefits. That is, despite the apparent contradiction that harming individuals shouldn't be a good evolutionary solution, senescence would actually be helping those individuals to  somehow increase the number of their descendants. Since this is an apparent contradiction, explanations and new models are needed to explain the gap between theory and data. 
A natural candidate to why individual selection could actually lead to an adaptation that can harm the individual interest is the concept of kin selection \cite{hamilton63a,hamilton64a,quellerstrassman98}. It might be in the best interest of the parents (in the sense of increasing the expected number of descendants) to die of senescence, leaving the available resources to their offspring \cite{bourke07a}. Spatial models \cite{travis04a,mitteldorf06a} have also been proposed but their results might be a consequence of kin selection effects. The introduction of diseases in a spatial model was also shown to have a positive influence into the adoption of senescence \cite{mitteldorfpepper09a}. And a model that ties senescence with the fact that organisms grow was also also proposed \cite{kaplanrobson09a}.

Kin selection could play an important role in the evolution of senescence. However, there is one part of this solution that sounds like circular reasoning. In cooperation problems, kin selection is actually a very good strategy because adult offspring is expected to produce larger progenies than their parents. This is particularly true when animals age. Parents are older, therefore, if senescence exists, they have an expected average number of children smaller than the number of their children. By transferring the resources to their offspring, they actually increase the expected number of their descendants. This reasoning fails when applied to senescence. If animals didn't age, there is no reason to assume that the children will be around longer so that they can produce more offspring than their parents. The expected gain will not be important unless there are mechanisms that gives an investment in the offspring a better expected return than keeping the resources for the parent. It might even be non-existent. Also, there are several results in the literature about cooperation pointing to the fact that while it makes sense for individuals to be altruistic to their relatives, if competition happens among those same relatives, the beneficial effects of kin selection can become smaller or disappear \cite{taylor92a,westeal02}. While there are conditions where kin selection can have an impact on the competition \cite{taylorirwin00a,lehmannrousset10a}, we will see evidence on the simulations that this is not the case in the models presented here.

I will show that this new conundrum can be solved if one notices a characteristic of the recent models, as they have some characteristics in common. In spatial models, conditions are different in each site and change over time. Diseases come and go, also introducing some non-stationarity characteristics into the problem. In the real world, change is actually much more common than in most evolutionary models. The environment changes, other species evolve, mutations happen. The conditions our ancestors were well adapted to are not necessarily the same ones as the conditions we have to live in. 

In this paper, I will present a model where two initially identical species compete for supremacy in a spatial grid. The only difference between them, at first, is that one of them dies of senescence, while the organisms of the second species could, in principle, live forever, if no accidents or competition happened. Each individual will be characterized by its ability to survive one more time step and, at each time step, alive organisms can produce offspring. In order to reflect the changing in the environment without actually having to worry about specifics, the fitness function, representing the survival capability of an individual, will decrease at each time step by a small amount. Also, offspring will not be an exact copy of their parents and, as such, their fitness can the the same, worse or better. I will show that when there is change, survival of an aging species can be the chosen by individual competition.

\section{A Model for Change}

 In the model presented here, individuals will compete in a landscape representing the world where they live. The environment will be represented by a square two-dimensional grid with $L^2$ sites, so that each individual will live and compete for the resources in one of the sites. Periodic boundary conditions will be assumed, so that no boundary effects are observed. Each site will have a carrying capacity $c=1$, that is, it can only sustain one individual at a time. Whenever more than $c=1$ individuals share the same site, they will compete for the local resources and only one will survive. Time will be measured in discrete steps, each time step corresponding to one generation, that is, the time the organisms need to produce new viable offspring. 
New offspring does not represent all the children of one individual, since only individuals who are at reproductive age are modeled. As such, if one species has many children and most of them die before reaching maturity, only the surviving child is described in the model and all others are assumed to have died between time steps. 

While most traditional evolutionary models work with a stationary environment as basis, which can be a very good first approximation, real world conditions are not unchanging. Climatic cycles happen, predators and prey evolve together in constant evolution, new diseases appear and replace old ones. Trying to model all those aspects and how they change with time would be a daunting task, with too many yet unsolved questions. Instead of doing that, an approximation will be adopted here. This will be implemented by proposing a type of fitness function, that captures the influence of the environment and the changing conditions. Unlike fitness functions of Evolutionary Game Theory \cite{maynardsmith82a,SigmundBook,weibull97a}, the one we will use here is not exactly the final payoff of a game the individuals play. But it plays a similar role, as it represents the likelihood an individual will survive sharing resources with a competitor.

Let $f_i(t)\ge 0$ be such a fitness function, that measures the likelihood that individual $i$ will defeat its competitors at time $t$. That means that, whenever there is competition between two (or more) agents in one site, the probability that agent $i$ will survive is proportional to $f_i(t)$. Given agent $i$ and agent $j$ competing for the resources in a site at time $t$, agent $i$ will survive with probability $f_i(t)/(f_i(t)+f_j(t))$. The larger $f_i$ is, the more likely $i$ is to survive, but there will always be a chance that less fit individuals would survive (except, of course, when $f_i=0$).

At each time step, surviving individuals produce offspring. Each offspring is born at a distance $b$ from its parent, where $b$ is measured in units of the grid size and it inherits the fitness of its parents, except for small deviations, due to mutation. For simplicity, if organism $i$ is the parent of organism $j$, the fitness of $j$ will be $f_j=f_i+m$, where $m=0,\pm M$, with equal probabilities for the three possible results. This represents small changes. Cases where rare, large and usually detrimental mutations happen will not be included in the model here, as strong detrimental mutations would almost certainly not survive until adulthood. The mutation here only represents the fact that surviving offspring can be a little different from their parents . Also, in order to represent the fact that conditions change, $f$ is diminished of a constant value every time step, so that $f_i(t+1)=f_i(t)-d$, where $d\ge 0$ for every individual $i$. 

Since the objective is to understand if aging related deaths can be chosen by an evolutionary process, two types of animals are introduced, those who die of senescence and those who will only die due to other causes. At first, both groups will always start with the same values for their parameters. Another approximation that will be made here is that all organisms who suffer the effects of senescence will die at the same age, $o$. 
In order to introduce the possibility of random deaths, disassociated from the competition, a chance $p_d$ that each organism will die at each time step can also be easily introduced. Tests with different values of $p_d$ showed little difference in the final chances of extinction for each species. Therefore, this will not be further explored in this paper.

\subsection{Aging and Competition} 

It is, of course, fundamental that the model, when no changes happen, should reproduce  some basic results of evolutionary aging theory. First, when the system is completely stable, no mutation going on and  no  changing conditions for worse, that is $M=d=0$, its is to be expected that 
a population that shows senescence will be driven to extinction. This happens simply because its members will die faster. And this is indeed the case. As a test, 20 runs were performed on a $51\times 51$ bi-dimensional grid, using NetLogo as platform \cite{wilensky99a}, where the species with senescence just survived until $o=5$ time steps. Offspring was born at a distance $b=1$. The species with senescence always died after 220-230 time steps. While this is not immediate, each time step corresponding to a new generation being born, the initial decline was still fast and the end, unavoidable.

That happens despite the fact that competition did cause old animals to be rare, both between the senescent and non-senescent species, with only approximately half of the individuals surviving longer than one time step.
Despite the small numbers of elder individuals, the evolutionary forces are strong and more than enough to prevent any possibility of survival for the aging species. Different values of the born distance provide the same scenario, with just different time scales for the extinction of the senescent species to happen. Per example, if diffusion is really slow, with $b=0.2$, it takes about 1,800 generations, in average, for the aging species to die out. 

\subsection{Mutation}

More interesting effects appear when change is introduced. Even with a fixed environment and change brought only as mutation, we can already observe new effects. The mutation introduces a random element that affects the competition between the species, with the obvious possibility that one of the species will get a better fitness $f$ by simple random chance. This should have a similar impact on both species and, at first sight, should not cause any qualitative differences from the results where no mutation happened. 

However, what we see is that there is a clear tendency for the aging species to adapt faster, meaning that $f$ has a strong tendency to be larger for the species that experiences senescence. This adaptability somehow compensates part of the detrimental effect that death by senescence has in the species. In a number of runs, it was observed that, for a while, the tendency to extinction of the senescent species was even reversed and its population increased in size, when the difference between average values of $f$ was large enough. Even tough senescence still caused the extinction of the aging species in most runs, it took longer for extinction to happen. And, even more important, there were a few cases where the aging group actually led the non-aging group to extinction!

As we increase the effects of mutation by making $M$ larger, per example, $M=0.1$ (and born distance $b=1$), we see that the non-senescent species finds it more difficult to drive the senescent one to extinction. While we observed extinction after an average of a little more than 200 generations when no mutation happened, the new average time to extinction grows to about 1,000 generations. Also, instead of a simple massacre where the non-aging species wins in every run, we observed that 7 out of 50 runs ended with the non-aging group extinct. Notice that this does not seem to be a kin selection effect, as the descendants of the non-aging group are exactly as important to the survival of the species as their parents. Even when $b=2$ and it is guaranteed that the parents will never compete with their direct offspring, 5 out of 50 runs ended with the aging group as the only survivors. An even stronger evidence against the fact the kin selection is responsible here is that fact that, when $b=0.5$, meaning that quite often parents will compete with their offspring, there was not a single run out of 50 where the aging species won at the end. Kin selection effects ought to be stronger the more the parents would compete with their offspring.

The eventual victory of the aging species in those few cases is almost certainly due to size effects and random fluctuation. And, indeed, when running a larger system with 101x101 sites, no victory of the aging species was observed in 20 realizations. However, the fact that finite size effects become more important shows that mutation makes aging more competitive. The increase in the competition brought by mutation is an important force in the system. And, in this case, the faster increase in the fitness function for aging species can actually make an important difference.

\subsection{Environmental changes}

  Introducing the idea that fitness decreases with time takes the model one more step closer to a more correct description of changes in the real world. As described above, this can be implemented by decreasing all $f$s by a constant amount, $d$, at each time step (generation). It is obviously an approximation to consider that the changes will affect all individuals equally and more random effects should be studied in the future. With the change in the environment, some degree of mutation is crucial, unless all $f$s will decline to zero fast. Also, if the mutation were too weak, selection forces wouldn't be able to keep up with the change in the environment. While this has minor consequences (up to the point where $f$ would become zero) in the model, as only the two species compete for the resources, in the real world, with more competitors, that could mean extinction for everyone.

Introduction of a tendency for the fitness to decrease ($d=0.005$) with time in the case described above($M=0.1$ and $b=1$) led to just a small increase in the number of victories by the aging group (10 in 50, instead of 7 in 50).  By making the change of the system even faster, $d=0.01$, we obtain a much more convincing case as almost half (23 in 50) of the runs ended with a victory for the aging species.
Investigating the possibility that these victories of the aging species can be due to finite size effects leads to a surprising result. When approximately doubling the length of the box, from a $51\times 51$ world to a $101\times 101$ (effectively allowing for four times the number of individuals in the system), the number of victories of the aging species actually increased to 39 in 50 runs. This actually suggests that now the survival of the non-aging species might be caused by the finite size.

\begin{figure}
\includegraphics[width=.75\textwidth]{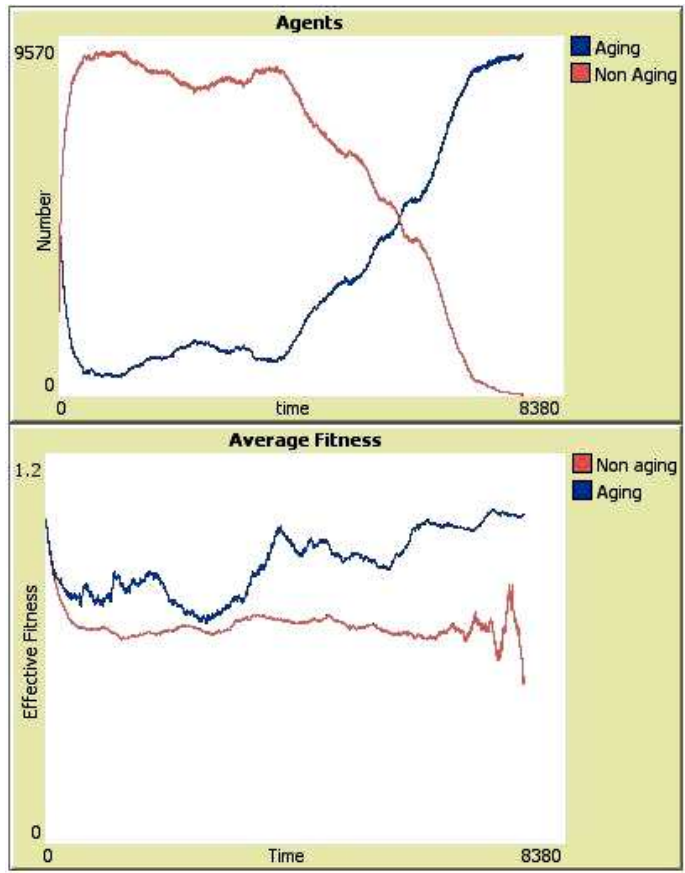}
\caption{The evolution of a typical run that ended with the victory of the senescent species. The blue lines correspond to the species where there is senescence (aging) and the red lines to the species where individuals do not age. {\it Top Panel}: Number of individuals (agents) in each group as a function of the number of generations (time). {\it Bottom Panel}: Evolution of the average value of the fitness function for each species.}
\label{Fig:aging}
\end{figure} 

In order to understand the mechanisms of survival under change, it is interesting to investigate how the system evolves with time. A typical run ending with a victory of the aging species can be seen in Figure \ref{Fig:aging}. It is easy to notice that the system is initially in a transient phase, due to non-natural initial conditions. When the simulation starts, an equal number of individuals of each species is created. Their location is randomly assigned and every agent has the same age of zero. As all spots are occupied soon after the start of the run, competition is unavoidable. The groups coalesce, with each species surviving in different areas, both as an effect of the competition and of chance. As expected, as the aging species starts losing a few of its members to senescence, the total number of aging individuals start to decrease.

\begin{figure}
\includegraphics[width=0.75\textwidth]{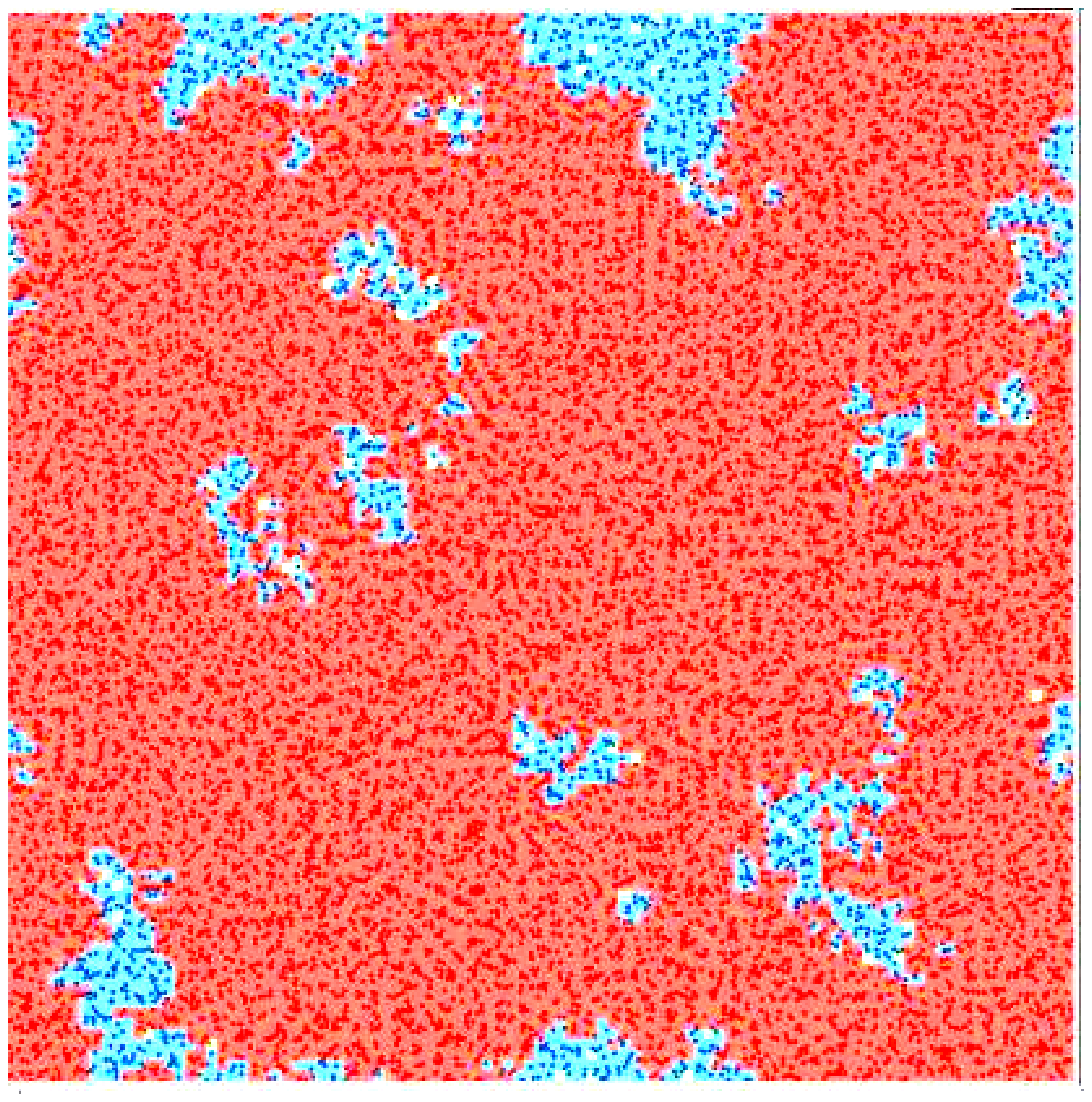}
\caption{Spatial distribution of aging and non-aging agents over the plots, after the initial diminishing in number of the aging agents. Aging agents are shown as blue dots and non-aging agents as red dots. In order to facilitate the visualization, the background of the square plot is also painted in light blue or light red, matching the color of the agent who lives there. The light green plots correspond to places where no agent is present, due to the fact that its inhabitant died of old age.}
\label{Fig:scatter}
\end{figure}

As the number of individuals in the aging species diminishes, with just small pockets scattered across the landscape (as can be seen in Figure \ref{Fig:scatter}), random variation becomes more important for the aging species, allowing its average fitness to oscillate faster. This can allow for a temporary recovery, as the fitness increases, but it also creates a larger chance for a deleterious change. If that was the whole story, mutation might be able to buy the aging species some time, as it does in the case with no change in the environment,  but the senescent species would eventually die.

But there is more. As the species start to compete, the pruning introduced by aging has a non trivial effect. It makes the pressure to adapt or perish stronger on the aging species; as its numbers dwindle, those who do survive are are fitter than those of the non-aging species. The reason for that is subtle. Due to mutation and the random selection of individuals, the average fitness of a new generation is a little larger than that of the previous ones. While it is possible for specific individuals to be lucky and survive for a while, on average, each generation is a little better adapted than the previous one. 

 Average fitness can be decreasing in time due to environment changes, as per example, in the beginning of the run shown in Figure \ref{Fig:aging}. But this affects every agent equally. When compared to the older generations, new generations have an average value of $f$ a little higher than the previous one.
That way, when senescence kills the elders, it eliminates a slightly worse adapted group. This leaves the space open for the newer, better adapted individuals and it increases their chance to survive. If this improvement is strong enough to compensate for the deaths from age, senescence will be chosen for its own evolutionary merits. 

Measurement of the difference between the average fitness of individuals with a difference in age of 4 time steps showed that the system tended to a point where the newer individuals indeed had a fitness that grew up until it was approximately 1\% larger. The same increasing pattern in the difference and same difference were observed in all 20 realizations where this difference was measured. Despite the small difference, the difference was consistent enough and, as such, not the effect of random variation.

Observing the evolution of each realization, we see that the average fitness of the senescent species becomes larger than that of the non senescent one and it stays larger. Fluctuations do happen, since this is a finite system, but the tendency is clear and it was repeated in all runs. Only exceptions where a reversal happened, with the non-aging species showing a higher fitness than the aging one, were due to large fluctuations when one of the two species had very small numbers. Thanks to this tendency to a larger fitness, if the difference in the fitness is large enough, the aging species can slowly drive off the non-aging one.

\begin{figure}
\includegraphics[width=.75\textwidth]{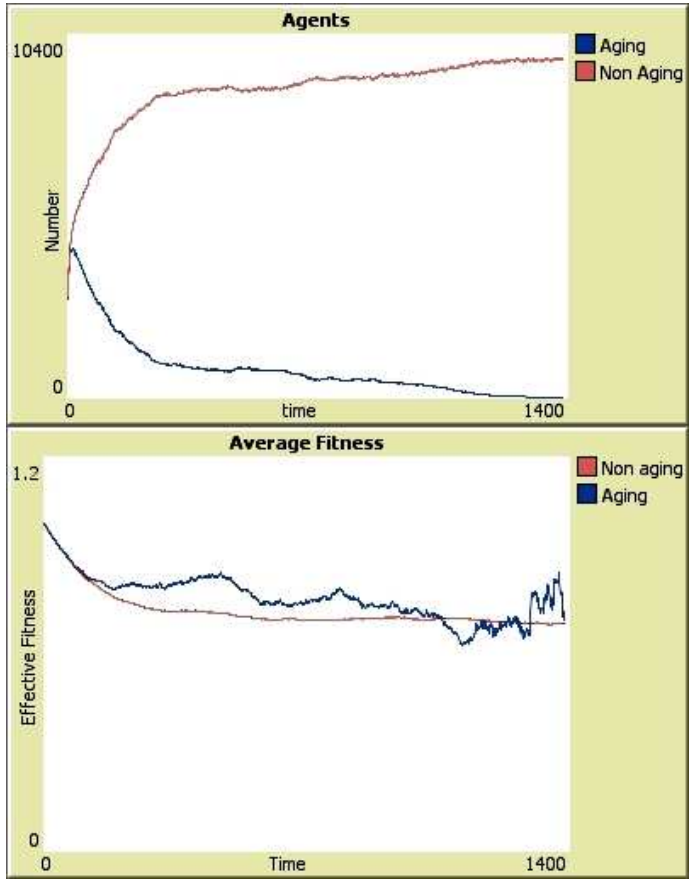}
\caption{The evolution of a typical run that ended with the defeat of the senescent species. The blue lines correspond to the species where there is senescence (aging) and the red lines to the species where individuals do not age. {\it Top Panel}: Number of individuals (agents) in each group as a function of the number of generations (time). {\it Bottom Panel}: Evolution of the average value of the fitness function for each species.}
\label{Fig:nonaging}
\end{figure}

In the cases where the immortal species drive the mortal one to extinction, dynamics is quite similar. A typical run ending with the victory of the non-aging species can be seen in Figure \ref{Fig:nonaging}. Both cases start in a basically identical way. Once all spaces are occupied and old age starts killing, the aging species start to decline. And, once again, its fitness becomes larger than the fitness of the non aging species. This allows the aging species to postpone extinction but, when a fluctuation in the average fitness make the advantage of the aging species smaller, the senescence starts taking its toll and the aging species is unable to recover from that oscillation.

When some amount of environment change is introduced, obviously, mutation becomes necessary to compensate for that. For a small amount of mutation, as soon as mutation becomes large enough to avoid a crash of the fitness function, the aging species actually shows a tendency to drive the non-aging species to extinction, according to the simulations performed for $d=0.002$, $0.004$, and $0.01$. For every value of $d$, it was observed that, as soon as the mutation $M$ was large enough to prevent a complete collapse of the fitness, it was the aging species that had the advantage of survival. For small changes in the environment ($d=0.002$), that change was not decisive and a maximum 65\% rate of success was observed. That rate declined to zero as $M$ got larger. For $0.004$, if mutation was just large enough to prevent the collapse, all 20 runs ended with the extinction of the non-aging species. The same effect was observed for faster changes in the environment ($d=0.01$).

For all values of the environment change $d$, however, as the variation $M$  associated with mutation became even larger, the advantage of the non-aging species declined until, for $M$ large enough, it disappeared. This meant that too strong variation actually led the aging species to extinction. This is actually to be expected. The survival of the aging species depended on the their average slightly better fitness. With strong variations, this advantage becomes too irregular, with both species able to gain a momentary better fitness.

\begin{figure}
\includegraphics[width=.95\textwidth]{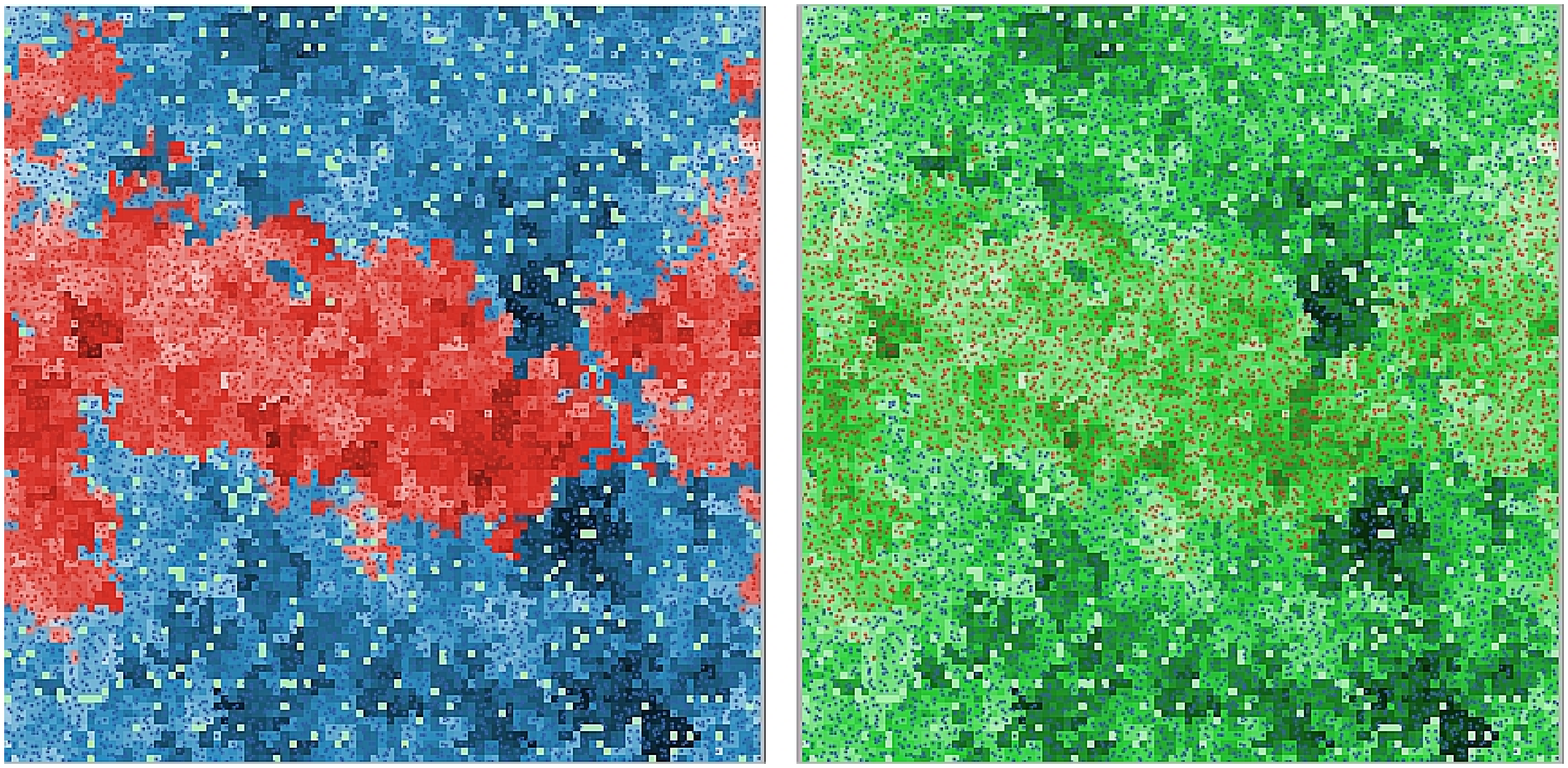}
\caption{Landscape of fitness values in the middle of a typical one run. These pictures correspond to a moment when the average fitness of the non aging species was 0.89 and that of the aging species, 1.17. {\it Left Panel} The division between the species is clearly shown, with red representing the non-aging species and blue, the aging species. Both for blue and red, lighter tones indicate a smaller fitness, while darker tones correspond to a higher fitness. {\it Right Panel} Same circumstances as in the left panel, but only fitness is shown in tones of green, with no distinction between the species, so that their fitness can be better compared. }
\label{Fig:fitness}
\end{figure}

Figure \ref{Fig:fitness} shows the spatial distribution of fitness in the middle of a typical run, when both species are still competing. The darker colors correspond to a higher fitness. The panel at the right, in green, shows the same moment, but with same green colors for both species, to allow a direct comparison. While a na\"ive guess would be that the agents with largest fit would be located at the interfaces, that is not the case and it is actually easy to understand why. From the agent point of view, others of the same species are as much a threat as those of a different species, as only one survives in each plot. That way, regions with higher fitness can be found both in the interfaces and inside the domains.

The right panel figure also shows clearly how the region that is occupied by the blue (aging) species corresponds to a region with average higher fitness than the red region (in that moment, the average fitness for the aging species was 1.17, and for the non-aging one, 0.89). The figures show that, not only the overall average is larger, but this tends to also happens in the frontiers. The larger fitness of the aging species when compared to their actual neighbors of the non-aging species allows its survival.The non-aging species would win, if both values were the same.

\section{Conclusion}

 Death by senescence certainly has an important evolutionary price for a species that adopt it. This is actually shown very clear in the model presented in this paper. This price is so obvious that it meant that senescence was considered for a long time as something that could not be an evolutionary selected characteristic. While the decrease in fitness makes for a strong argument, that this is not the whole story. By introducing gradual change in the environment, under the right circumstances, senescence can be chosen by evolutionary dynamics as the best answer to that change. Aging can produce a pruning effect on the species, eliminating older, slightly less adapted individuals who had managed to survive by chance. 

Chance and change were the fundamental keys to this answer. While a larger fitness ensures a better chance of survival, this is only a better probability, both in the model as well as in the real world. If the difference in the fitness between the individuals is not so big, it is to be expected that the better prepared wins only in average. When this is associated with random mutations and environmental changes, an aging species can have a small advantage to compensate for the deaths by old age and, in the long run, drive the non-aging species to extinction. A model based on selection of individuals can lead to an evolutionary advantage of a senescent species, when change is incorporated. This finally explains the apparent paradox of why we age. And it illustrates how we still don't understand all consequences of change and random chance in a system as complex as the natural world.

{\it \bf Acknowledgements} The author would like to thank Funda\c{c}\~ao de Amparo \`a Pesquisa do Estado de S\~ao Paulo (FAPESP) for partial support to this work, under grant 2009/08186-0.

\bibliographystyle{plain}
\bibliography{biblio}

\end{document}